\documentclass[a4paper,12pt]{article}

 \usepackage{amsfonts}
\usepackage{verbatim}

\begin{document}

\begin{flushright} arXiv: 0805.1102 (hep-th)\\ CAS-PHYS-BHU Preprint
\end{flushright} 

\vskip 1cm

\begin{center}

{\small \sf A NOTE ON THE (ANTI-)BRST INVARIANT LAGRANGIAN DENSITIES FOR
THE FREE ABELIAN 2-FORM GAUGE THEORY}\\

\vskip 2cm

{\small \sf SAURABH GUPTA$^{(a)}$, R. P. MALIK$^{(a, b)}$}\\
{\it $^{(a)}$Physics Department, Centre of Advanced Studies,}\\
{\it Banaras Hindu University, Varanasi - 221 005, (U.P.), India}\\

\vskip 0.1cm

{\bf and}\\

\vskip 0.1cm
{\it $^{(b)}$DST Centre for Interdisciplinary Mathematical Sciences,}\\
{\it Faculty of Science, Banaras Hindu University, Varanasi - 221 005, India}\\
{\small {\bf e-mails: guptasaurabh4u@gmail.com, malik@bhu.ac.in}}

\end{center}

\vskip 2.5cm

\noindent
{\bf Abstract:}
We show that the previously known off-shell nilpotent ($s_{(a)b}^2 = 0$) and absolutely anticommuting
($s_b s_{ab} + s_{ab} s_b = 0$) Becchi-Rouet-Stora-Tyutin (BRST) transformations ($s_b$) 
and anti-BRST transformations ($s_{ab}$) are the symmetry transformations of the 
appropriate Lagrangian densities of a four (3 + 1)-dimensional (4D) free Abelian 2-form gauge theory which do {\it not} explicitly incorporate a very specific constrained field condition 
through a Lagrange multiplier 4D vector field. The above condition, which is the analogue of the Curci-Ferrari restriction of the non-Abelian 1-form gauge theory, emerges from the Euler-Lagrange 
equations of motion of our present theory and ensures the 
{\it absolute} anticommutativity of the transformations $s_{(a)b}$. 
Thus, the coupled Lagrangian densities, proposed in our present investigation, are 
aesthetically more appealing and more economical.\\



\newpage

\section{Introduction}

The principle of local gauge invariance provides a precise theoretical basis for the description of the three (out of four) fundamental interactions of nature. The theories with local gauge symmetries are always (i) described by the singular Lagrangian densities, and (ii) endowed with the first-class constraints in the language of Dirac's prescription for the classification scheme [1,2]. It has been well-established that the latter (i.e. the first-class constraints) generate the above local gauge symmetry transformations for the singular Lagrangian densities of the relevant gauge theories.

One of the most attractive approaches to covariantly quantize such kind of theories is the BRST formalism where (i) the unitarity and ``quantum'' gauge (i.e. BRST)
invariance are respected together, (ii) the true physical states are defined in terms of the BRST charge which turn out to be consistent with the Dirac's prescription for the quantization of  systems with constraints, and (iii) there exists a deep relationship between the physics of the gauge theories (in the framework of BRST formalism) and the mathematics of differential geometry (e.g. cohomology) and supersymmetry (e.g. superfield formalism).

Some of the key and cute mathematical properties, associated with the (anti-)BRST symmetry transformations, are as follows. First, there exist two symmetry transformations
(christened as the (anti-)BRST\footnote{We follow here the standard notations and conventions adopted in  our recent works on 4D free Abelian 2-form gauge theory within the framework
of BRST formalism [3-5].} symmetry transformations
$s_{(a)b}$) for a given
local gauge symmetry transformation. Second, both the symmetries are nilpotent of order two (i.e. $s_{(a)b}^2 = 0$). Finally, they anticommute
(i.e. $s_b s_{ab} + s_{ab} s_b = 0$) with each-other when they act {\it together} on any specific field of the theory. These properties are very sacrosanct for any arbitrary
gauge (or reparametrization) invariant theory when the latter
 is described within the framework of the BRST formalism.

Recently, the 2-form ($B^{(2)} = (1/2!) (dx^\mu \wedge dx^\nu) B_{\mu\nu}$) Abelian gauge field $B_{\mu\nu}$ [6,7] 
and corresponding gauge theory have attracted a great deal of interest because of their relevance in the context of (super)string theories.
This Abelian 2-form gauge theory has also been shown to provide (i) an explicit field theoretical example of the Hodge theory [8], and (ii) a model for the quasi-topological field theory [9].
The (anti-)BRST invariant Lagrangian densities of the 2-form 
theory have been written out and
the BRST quantization has been performed [8-11]. One of the key observations in (see, e.g. [8,9])
is that the above (anti-)BRST transformations, even though precisely off-shell nilpotent, are 
found to be anticommuting 
only up to a vector gauge transformation. Thus, the absolute anticommutativity property is lost.

As pointed out earlier, the anticommutativity property of the (anti-)BRST symmetry transformations is a cardinal requirement in the domain of application of the BRST formalism to gauge theories. This key property actually encodes the linear independence of the above two transformations corresponding to a given local gauge symmetry
transformation (of a specific gauge theory). In the realm of superfield approach to BRST formalism (see, e.g.,
[12,5]), the absolute anticommutativity of these transformations 
becomes crystal clear because these 
are identified with the translational generators along the Grassmannian directions of a
(D, 2)-dimensional supermanifold on which any arbitrary D-dimensional gauge theory is considered [12,5].

It is worthwhile to mention that, the superfield approach, proposed in [12] for the 4D non-Abelian 1-form
gauge theory, has been applied, for the first time, to the description of the free 4D Abelian 2-form
gauge theory in [5]. One of the upshots (of the discussions in [5]) is that an analogue of
the Curci-Ferrari (CF) type of restriction [13] emerges in the context of 4D {\it Abelian} 2-form gauge theory.
The former happens to be the hallmark of a 4D {\it non-Abelian} 1-form gauge theory [12,13]. This CF type condition
ensures (i) the absolute anticommutativity of the (anti-)BRST symmetry transformations of the Abelian
2-form gauge theory, and (ii) the identification of the (anti-) BRST symmetry transformations
with the translational generators along the Grassmannian directions of the (4, 2)-dimensional
supermanifold [5].

Keeping the above properties in mind, the 
(anti-)BRST symmetry transformations have been obtained in our earlier works [3,4] where the above
CF type field condition is invoked for the proof of the absolute
anticommutativity of the off-shell nilpotent (anti-)BRST symmetry transformations [3,4]. In fact,
the above field condition is explicitly incorporated in the Lagrangian densities through a Lagrange
multiplier vector field (which is not a basic dynamical field of the theory). Furthermore, due to the above
restriction, the kinetic term for the massless scalar field of the theory 
turns out to possess a negative sign. 
These are the prices one pays to obtain the absolute anticommutativity of the 
nilpotent (anti-)BRST symmetry transformations.

The purpose of our present investigation is to show that the 
(anti-)BRST tansformations of our earlier works
[3,4] are the {\it symmetry} transformations of a pair of coupled Lagrangian densities which do
not incorporate the analogue of the CF type restriction explicitly through the Lagrange multiplier
4D vector field\footnote{This feature is exactly like the discussion of the absolutely anticommuting
(anti-)BRST symmetry transformations in the context of the 4D non-Abelian 1-form gauge theory.}. This condition, however, appears in the theory as a consequence of the Euler-Lagrange equations of motion that are derived 
from the coupled Lagrangian densities.
Furthermore, all the terms of these Lagrangian densities carry standard meaning and there are 
no peculiar signs associated with any of them. One of the key features of the CF type restriction, for our present Abelian theory, is that it does not involve any kind of (anti-)ghost fields. 
On the contrary, one knows that the original CF restriction of the non-Abelian 1-form gauge theory [13] does involve the (anti-)ghost fields.

The key factors that have propelled us to pursue our present investigation are as follows. First and foremost, it is very important to obtain the correct nilpotent and anticommuting (anti-)BRST symmetry transformations which are respected by the appropriate Lagrangian densities. The latter 
should, for aesthetic reasons, be economical and beautiful (i.e. possessing no peculiar looking terms). Second, the theory itself should produce all the cardinal requirements and nothing should be imposed from outside through a Lagrange multiplier field. Third, the (anti-)BRST symmetry transformations in the Lagrangian formulation [3,4] must be  consistent with the derivation of the same from the superfield approach [5]. Finally, our present study is the first modest step towards our main goal of applying the BRST formalism to higher p-form ($p > 2$) gauge theories
that are relevant in (super)string theories.

The contents of our present investigation are organized as follows.
In Sec. 2, we briefly recapitulate the bare essentials of the off-shell
nilpotent and anticommuting (anti-)BRST symmetry transformations for a couple of
Lagrangian densities of the Abelian 2-form gauge theory.
The above Lagrangian densities incorporate a constrained field relationship
through a Lagrange multiplier 4D vector field. Our Sec. 3 deals with a pair of coupled
and equivalent Lagrangian densities that (i) respect the BRST and anti-BRST symmetry
transformations, and (ii) do not incorporate any constrained field relationship explicitly.
In Sec. 4, we derive an explicit BRST algebra by exploiting the 
infinitesimal continuous symmetry transformations. 
We make some concluding remarks in our Sec. 5.


\section{Preliminaries: Lagrangian Densities Incorporating the Constrained Field Condition}

\noindent
We begin with the following nilpotent (anti-)BRST 
symmetry invariant Lagrangian density for the 4D
free Abelian 2-form gauge theory
\footnote{We choose the 4D spacetime metric $\eta_{\mu\nu}$ with the signatures
$(+ 1, - 1, -1, -1)$ so that $P \cdot Q = \eta_{\mu\nu} P^\mu Q^\nu = P_0 Q_0 - P_i Q_i$
is the dot product between non-null
four vectors $P_\mu$ and $Q_\mu$. Here $\mu, \nu, \kappa, \sigma...= 0 , 1, 2, 3$ and
$i, j, k.... = 1, 2, 3$. We also adopt, in the whole body of our text, the 
field differentiation convention:  
$(\delta B_{\mu\nu}/\delta B_{\kappa\sigma})
= \frac{1}{2!} (\delta_{\mu\kappa} \delta_{\nu\sigma} 
- \delta_{\mu\sigma} \delta_{\nu\kappa})$, etc.} [3-5]
\begin{eqnarray}
{\cal L}^{(1)} &=& \frac{1}{6} H^{\mu\nu\kappa} H_{\mu\nu\kappa} + B^\mu (\partial^\nu B_{\nu\mu}) + \frac{1}{2} (B \cdot B + \bar B \cdot \bar B)
- \frac{1}{2} \partial_\mu \phi\partial^\mu \phi \nonumber\\
&+&  \partial_\mu \bar\beta \partial^\mu \beta
+ (\partial_\mu \bar C_\nu - \partial_\nu \bar C_\mu) (\partial^\mu C^\nu) +
(\partial \cdot C - \lambda) \rho \nonumber\\
&+& (\partial \cdot \bar C + \rho) \lambda + L^\mu (B_\mu - \bar B_\mu - \partial_\mu \phi),
\end{eqnarray}
where the kinetic term is constructed with the totally antisymmetric curvature tensor
$H_{\mu\nu\kappa}$ which is derived from the 3-form $H^{(3)} = \frac{1}{3!} (dx^\mu \wedge dx^\nu \wedge dx^\kappa) H_{\mu\nu\kappa}$. The exterior derivative $d = dx^\mu \partial_\mu$ 
(with $d^2 = 0$) and the 2-form $B^{(2)} = \frac{1}{2!} (dx^\mu \wedge dx^\nu) B_{\mu\nu}$
generate the above 3-form (i.e. $H^{(3)} = d B^{(2)}$).

We have the Lorentz
vector fermionic (anti-)ghost
fields $(\bar C_\mu)C_\mu$ and the bosonic (anti-)ghost fields $(\bar \beta)\beta$ in the theory.
The above Lagrangian density also requires fermionic auxiliary ghost fields $\rho = - \frac{1}{2}
(\partial \cdot \bar C)$ and $\lambda = \frac{1}{2} (\partial \cdot C)$. The auxiliary vector
fields $B_\mu$ and $\bar B_\mu$ are constrained to satisfy the field equation $B_\mu - \bar B_\mu - \partial_\mu \phi = 0$ where the massless (i.e. $\Box \phi = 0$) field $\phi$ is required for
the stage-one reducibility in the theory. The above constrained field equation emerges due
to presence of the Lagrange multiplier field $L^\mu$.

The following off-shell nilpotent (i.e. $s_{b}^2 = 0$) BRST symmetry transformations $s_{b}$
for the 4D local fields of the theory, namely;
\begin{eqnarray}
&& s_b B_{\mu\nu} = - (\partial_\mu C_\nu - \partial_\nu C_\mu), \qquad s_b C_\mu = - \partial_\mu \beta,
 \qquad s_b \bar C_\mu = - B_\mu, \nonumber\\
&& s_b L_\mu = - \partial_\mu \lambda,\; \qquad
s_b \phi = \lambda,\; \qquad s_b \bar \beta = - \rho, \nonumber\\
&& s_b \bar B_\mu = - \partial_\mu \lambda,\; \;\qquad\;
s_b \bigl [\rho, \lambda, \beta, B_\mu, H_{\mu\nu\kappa} \bigr ] = 0,
\end{eqnarray}
leave the above Lagrangian density quasi-invariant because it transforms to a total spacetime derivative: $s_b {\cal L}^{(1)} = - \partial_\mu [(\partial^\mu C^\nu - \partial^\nu C^\mu) B_\nu + \lambda B^\mu + \rho \partial^\mu \beta  ]$.

In exactly similar fashion, the following off-shell nilpotent ($s_{ab}^2 = 0$) anti-BRST
symmetry transformations $s_{ab}$
\begin{eqnarray}
&& s_{ab} B_{\mu\nu} = - (\partial_\mu \bar C_\nu - \partial_\nu
\bar C_\mu),\; \quad s_{ab} \bar C_\mu = - \partial_\mu \bar \beta,
\quad s_{ab}  C_\mu = + \bar B_\mu, \nonumber\\
&&  s_{ab} L_\mu = - \partial_\mu \rho, 
\qquad 
s_{ab} \phi = \rho,\; \qquad s_{ab} \beta = - \lambda, \nonumber\\
&&  s_{ab} B_\mu = +
\partial_\mu \rho, \qquad  s_{ab} \bigl [\rho, \lambda, \bar\beta,
\bar B_\mu, H_{\mu\nu\kappa} \bigr ] = 0,
\end{eqnarray}
leave the following Lagrangian density
\begin{eqnarray}
{\cal L}^{(2)} &=& \frac{1}{6} H^{\mu\nu\kappa} H_{\mu\nu\kappa} + \bar B^\mu 
(\partial^\nu B_{\nu\mu}) + \frac{1}{2} (B \cdot B + \bar B \cdot \bar B)
- \frac{1}{2} \partial_\mu \phi\partial^\mu \phi \nonumber\\
&+&  \partial_\mu \bar\beta \partial^\mu \beta
+ (\partial_\mu \bar C_\nu - \partial_\nu \bar C_\mu) (\partial^\mu C^\nu) +
(\partial \cdot C - \lambda) \rho \nonumber\\
&+& (\partial \cdot \bar C + \rho) \lambda + L^\mu (B_\mu - \bar B_\mu - \partial_\mu \phi),
\end{eqnarray}
quasi-invariant because it transforms to a total spacetime derivative as is
evident from $s_{ab} {\cal L}^{(2)} = - \partial_\mu  [(\partial^\mu \bar C^\nu 
- \partial^\nu \bar C^\mu) \bar B_\nu
- \rho \bar B^\mu + \lambda \partial^\mu \bar \beta ]$. It is interesting
to point out that both the Lagrangian densities (1) and (4) respect the
off-shell nilpotent (anti-)BRST
symmetry transformations (cf. (2) and (3)) on a constrained surface defined by a field
equation (see, e.g. equation (5) below).

Both the above nilpotent transformations $s_{(a)b}$ (cf. (2) and (3))
are absolutely {\it anticommuting} (i.e. $s_b s_{ab} + s_{ab} s_b \equiv \{s_b, s_{ab} \} = 0$)
in nature if the whole 4D free Abelian 2-form gauge theory is defined on 
a constrained surface parametrized by the following field equation
\footnote{This restriction comes out from our
previous work [5] that is devoted to the discussion 
of the free 4D Abelian 2-form gauge theory within the framework of
superfield formalism.} 
\begin{equation}
B_\mu - \bar B_\mu -\partial_\mu \phi = 0.
\end{equation}
This is due to the fact that  $\{ s_b, s_{ab} \} B_{\mu\nu} = 0$ 
is true only if the above equation is satisfied. This
condition has been incorporated in the above Lagrangian densities through the
Lagrange multiplier Lorentz 4D vector field $L^\mu$.

The Lagrangian densities (1) and (4) are coupled Lagrangian densities on the constrained
field surface defined by (5). It would be very nice if one could obtain Lagrangian densities
that respect the nilpotent and anticommuting 
(anti-)BRST symmetry transformations (2) and (3) and are free of
any Lagrange multiplier field. The latter fields are required when we wish to put some
restriction, from outside, on the theory. A beautiful theory 
should produce this restriction
on its own strength. Thus, it is desired that the Lagrangian density
of a theory should be devoid of Lagrange multipliers.
Furthermore, it would be better if we could avoid
the negative kinetic term for the massless scalar field $\phi$ that is present in 
the Lagrangian densities (1) and (4) of our present theory. We address these issues
in our next section.

\section{Lagrangian Densities Without Any Constrained Field Condition: Symmetries}

It is interesting to note that the following coupled and equivalent 
(cf. (5)) Lagrangian densities
for the  4D free Abelian 2-form gauge theory, namely;
\begin{eqnarray}
{\cal L}_B &=& \frac{1}{6} H^{\mu\nu\kappa} H_{\mu\nu\kappa} + B^\mu (\partial^\nu B_{\nu\mu} - \partial_\mu \phi) + B \cdot B + \partial_\mu \bar\beta \partial^\mu \beta\nonumber\\
&+& (\partial_\mu \bar C_\nu - \partial_\nu \bar C_\mu) (\partial^\mu C^\nu) +
(\partial \cdot C - \lambda) \rho + (\partial \cdot \bar C + \rho)\; \lambda,
\end{eqnarray}
\begin{eqnarray}
{\cal L}_{\bar B} &=& \frac{1}{6} H^{\mu\nu\kappa} H_{\mu\nu\kappa} + \bar B^\mu (\partial^\nu B_{\nu\mu} + \partial_\mu \phi) + \bar B \cdot \bar B + \partial_\mu \bar\beta \partial^\mu \beta\nonumber\\
&+& (\partial_\mu \bar C_\nu - \partial_\nu \bar C_\mu) (\partial^\mu C^\nu) +  (\partial \cdot C - \lambda) \rho + (\partial \cdot \bar C + \rho)\; \lambda,
\end{eqnarray}
remain quasi invariant under the nilpotent and anticommuting (anti-)BRST symmetry transformations (2) and (3), respectively. However, these Lagrangian densities do not incorporate explicitly the constrained field condition (5). Neither do they possess negative kinetic term for the massless scalar field $\phi$. Thus, above Lagrangian densities are the appropriate ones.

The above Lagrangian densities (6) and (7) are equivalent on the constrained surface (defined
by the field equation (5)) because they respect both the BRST and anti-BRST symmetry transformations
separately and independently.
To clarify this statement explicitly, it can be checked that the Lagrangian density (6)
transforms under the off-shell nilpotent (anti-)BRST symmetry transformations as given below 
\begin{eqnarray}
s_b {\cal L}_B &=& s_b {\cal L}^{(1)}, \nonumber\\
s_{ab} {\cal L}_B &=& - \partial_\mu [ (\partial^\mu \bar C^\nu - \partial^\nu \bar C^\mu) B_\nu + \lambda \partial^\mu \bar \beta \nonumber\\
&-& \rho (\partial_\nu B^{\nu\mu} + \bar B^\mu)] + (B^\mu - \bar B^\mu - \partial^\mu \phi) \partial_\mu \rho 
\nonumber\\
&+& \partial^\mu (B^\nu - \bar B^\nu - \partial^\nu \phi) (\partial_\mu \bar C_\nu - \partial_\nu \bar C_\mu).
\end{eqnarray}
In an exactly similar fashion, the Lagrangian density (7) changes under the (anti-)BRST
symmetry transformations as
\begin{eqnarray}
s_{ab} {\cal L}_{\bar B} &=& s_{ab} {\cal L}^{(2)}, \nonumber\\
s_{b} {\cal L}_{\bar B} &=& - \partial_\mu [ (\partial^\mu  C^\nu - \partial^\nu  C^\mu) \bar B_\nu + \rho \partial^\mu  \beta \nonumber\\
&+& \lambda (\partial_\nu B^{\nu\mu} +  B^\mu)] + (B^\mu - \bar B^\mu - \partial^\mu \phi) \partial_\mu \lambda 
\nonumber\\
&-& \partial^\mu (B^\nu - \bar B^\nu - \partial^\nu \phi) (\partial_\mu  C_\nu - \partial_\nu  C_\mu).
\end{eqnarray}
Thus, on the constrained surface (defined by (5)), the
Lagrangian densities (6) and (7) are equivalent
and both of them respect the (anti-)BRST symmetry invariances. The
condition (5), however, has to be imposed from outside.

The following Euler-Lagrange equations of motion 
\begin{eqnarray}
B_\mu = - \frac{1}{2} (\partial^\nu B_{\nu\mu} - \partial_\mu \phi), \qquad
\bar B_\mu = - \frac{1}{2} (\partial^\nu B_{\nu\mu} + \partial_\mu \phi),
\end{eqnarray}
from the above Lagrangian densities (6) and (7) imply that
\begin{eqnarray}
&& \partial \cdot B = 0,\; \qquad \partial \cdot \bar B = 0,\; \qquad \Box \phi = 0, \nonumber\\
&& B_\mu - \bar B_\mu - \partial_\mu \phi = 0,\; \qquad
B_\mu + \bar B_\mu + \partial^\nu B_{\nu\mu} = 0.
\end{eqnarray}
Thus, the analogue of the Curci-Ferrari restriction [13] of the non-Abelian 1-form gauge theory,
is hidden in the above coupled Lagrangian densities in the form of the Euler-Lagrange equation
of motion (cf. (11) {\it vis-{\`a}-vis} (5)).

To capture the above (anti-)BRST invariance in a simpler setting, it can be seen that 
the Lagrangian densities (6) and (7) can be re-expressed as the sum of the kinetic term
and the BRST and anti-BRST exact forms, namely;
\begin{equation}
{\cal L}_B = \frac{1}{6} H^{\mu\nu\kappa} H_{\mu\nu\kappa} + s_b \Bigl [ - \bar C^\mu \bigl \{ (\partial^\nu B_{\nu\mu} - \partial_\mu \phi)
+ B_\mu \bigr \} + \bar \beta \bigl (\partial \cdot C - 2 \lambda \bigr ) \Bigr ],
\end{equation}
\begin{equation}
{\cal L}_{\bar B} = \frac{1}{6} H^{\mu\nu\kappa} H_{\mu\nu\kappa} +
s_{ab} \Bigl [ + C^\mu \bigl \{ (\partial^\nu B_{\nu\mu} + \partial_\mu \phi)
+ \bar B_\mu \bigr \} + \beta \bigl (\partial \cdot \bar C + 2 \rho \bigr ) \Bigr ].
\end{equation}
The above equations provide a simple and straightforward proof for the nilpotent symmetry invariance
of the Lagrangian densities (6) and (7) because of (i) the nilpotency (i.e. $ s_{(a)b}^2 = 0$)
of the transformations $s_{(a)b}$, and (ii) the invariance of the curvature term
(i.e. $s_{(a)b} H_{\mu\nu\kappa} = 0$) under $s_{(a)b}$.

It will be noted that the following interesting expressions\footnote{These
relations are similar to the case of non-Abelian 1-form gauge theory where
the CF restriction is {\it not} explicitly incorporated in the Lagrangian densities (see, e.g., [12]).}
\begin{eqnarray}
&&s_b s_{ab} \Bigl [ 2 \beta \bar\beta + \bar C_\mu C^\mu - \frac{1}{4}
B^{\mu\nu} B_{\mu\nu} \Bigr ] =
 B^\mu (\partial^\nu B_{\nu\mu})  + B \cdot \bar B
+ \partial_\mu \bar\beta \partial^\mu \beta \nonumber\\
&& + (\partial_\mu \bar C_\nu - \partial_\nu \bar C_\mu) (\partial^\mu C^\nu) +
(\partial \cdot C - \lambda) \rho + (\partial \cdot \bar C + \rho) \lambda,
\end{eqnarray}
\begin{eqnarray}
&&- s_{ab} s_b \Bigl [ 2 \beta \bar\beta + \bar C_\mu C^\mu - \frac{1}{4}
B^{\mu\nu} B_{\mu\nu} \Bigr ] =
 \bar B^\mu (\partial^\nu B_{\nu\mu}) + B \cdot \bar B
+ \partial_\mu \bar\beta \partial^\mu \beta \nonumber\\
&& + (\partial_\mu \bar C_\nu - \partial_\nu \bar C_\mu) (\partial^\mu C^\nu) +
(\partial \cdot C - \lambda) \rho + (\partial \cdot \bar C + \rho) \lambda,
\end{eqnarray}
allow us to express the Lagrangian densities (6) and (7) in yet another forms
\begin{equation}
{\cal L}_B = \frac{1}{6} H^{\mu\nu\kappa} H_{\mu\nu\kappa} + 
s_b s_{ab} \Bigl [ 2 \beta \bar\beta + \bar C_\mu C^\mu - \frac{1}{4}
B^{\mu\nu} B_{\mu\nu} \Bigr ],
\end{equation}
\begin{equation}
{\cal L}_{\bar B} = \frac{1}{6} H^{\mu\nu\kappa} H_{\mu\nu\kappa} 
- s_{ab} s_b \Bigl [ 2 \beta \bar\beta + \bar C_\mu C^\mu - \frac{1}{4}
B^{\mu\nu} B_{\mu\nu} \Bigr ],
\end{equation}
where one has to make use of (5) (or (11)) to express $(B \cdot \bar B)$
either equal to ($B \cdot B - B^\mu \partial_\mu \phi$) or equal
to  ($\bar B \cdot \bar B + \bar B^\mu \partial_\mu \phi$). Once again,
one can note the (anti-)BRST invariance of the Lagrangian densities (17)
and (16) due to the nilpotency ($ s_{(a)b}^2 = 0$) and invariance of the curvature term
($s_{(a)b} H_{\mu\nu\kappa} = 0$). It is worthwhile to mention that the
Lagrangian densities in (1) and (4) cannot be recast into the forms
like equations (12), (13), (16) and (17). The central obstacle in
this attempt 
is created by the Lagrange multiplier term and kinetic term
for the massless scalar field $\phi$ (cf. (1) and (4)).

The following global transformations of the fields 
\begin{eqnarray}
&&B_{\mu\nu} \to B_{\mu\nu}, \qquad B_\mu \to B_\mu, \qquad \bar B_\mu \to \bar B_\mu,
\qquad \phi \to \phi, \nonumber\\
&& \beta \to e^{+ 2 \Omega} \beta,\; \qquad \bar \beta \to e^{- 2 \Omega} \bar\beta,\; \qquad
C_\mu \to e^{+ \Omega} C_\mu, \nonumber\\ && \bar C_\mu \to e^{- \Omega} \bar C_\mu,\; \qquad
\lambda \to e^{+ \Omega} \lambda,\; \qquad \rho \to e^{-\Omega} \rho,
\end{eqnarray}
(where $\Omega$ is an infinitesimal  global parameter) leave the Lagrangian densities (6) and (7)
invariant. A close look at the above transformations
shows that all the ghost terms of (6) and (7) 
remain invariant under the above transformations. The
infinitesimal version of the above global ghost transformations $s_g$ 
(modulo parameter $\Omega$) is
such that $s_g \beta = 2 \beta, s_g \bar \beta = - 2 \bar\beta, 
s_g C_\mu = + C_\mu, s_g \bar C_\mu = -  \bar C_\mu, s_g \lambda = + \lambda, 
s_g \rho = - \rho$.
The factors of $\pm 2$ and $\pm 1$, present in the exponentials of equation (18),
correspond to the ghost numbers of the corresponding ghost fields which would play very 
significant roles in the next section where we shall compute some commutators with
the ghost charge.

\section{Generators of the Continuous Symmetry \\Transformations: BRST Algebra}

The nilpotent (anti-)BRST symmetry transformations (3) and (2) and the infinitesimal
version of the global transformations in (18) lead to the derivation of the Noether conserved
currents. These are as follows
\begin{eqnarray}
J^\mu_{(ab)} &=&  \rho \bar B^\mu
-(\partial^\mu  C^\nu - \partial^\nu  C^\mu) \partial_\nu \bar \beta
- H^{\mu\nu\kappa} (\partial_\nu \bar C_\kappa - \partial_\kappa \bar C_\nu) \nonumber\\
&-& \lambda \partial^\mu \bar \beta  - (\partial^\mu \bar C^\nu 
- \partial^\nu \bar C^\mu) \bar B_\nu, \nonumber\\
J^\mu_{(b)} &=& (\partial^\mu \bar C^\nu - \partial^\nu \bar C^\mu) \partial_\nu \beta
- H^{\mu\nu\kappa} (\partial_\nu C_\kappa - \partial_\kappa C_\nu) \nonumber\\
&-& \rho \partial^\mu \beta - \lambda B^\mu - (\partial^\mu C^\nu - \partial^\nu C^\mu) B_\nu,
\nonumber\\
J^\mu_{(g)} &=& 2 \beta \partial^\mu \bar\beta - 2 \bar\beta \partial^\mu \beta 
+ \lambda \bar C^\mu - \rho C^\mu \nonumber\\
&+& (\partial^\mu \bar C^\nu - \partial^\nu \bar C^\mu) C_\nu + 
(\partial^\mu  C^\nu - \partial^\nu  C^\mu) \bar C_\nu.
\end{eqnarray}
It is straightforward to check that the continuity equation
$\partial_\mu J^\mu_{(i)} = 0$ (with $ i = b, ab, g$) is satisfied if we exploit the
Euler-Lagrange equations of motion derived from the Lagrangian densities (6) and (7).

The above Noether conserved currents lead to the definition of the conserved
and nilpotent
($Q^2_{(a)b} = 0$) (anti-)BRST charges ($Q_{(a)b} = \int d^3 x J^0_{(a)b}$) 
and the conserved ghost charge ($Q_g = \int d^3 x J^0_{(g)}$)
as given below
\begin{eqnarray}
Q_{ab} &=& {\displaystyle \int} d^3 x \Bigl [ \rho \bar B^0 - \lambda \partial^0 \bar \beta
- H^{0ij} (\partial_i \bar C_j - \partial_j \bar C_i) \nonumber\\
&-& (\partial^0 \bar C^i - \partial^i \bar C^0) \bar B_i - 
(\partial^0  C^i - \partial^i  C^0) \partial_i \bar \beta \Bigr ], \nonumber\\
Q_{b} &=& {\displaystyle \int} d^3 x \Bigl [ 
(\partial^0  \bar C^i - \partial^i \bar C^0) \partial_i  \beta
- H^{0ij} (\partial_i  C_j - \partial_j  C_i) \nonumber\\
&-& (\partial^0  C^i - \partial^i  C^0)  B_i - \lambda B^0 - \rho \partial^0 \beta
 \Bigr ], \nonumber\\
Q_{g} &=&  {\displaystyle \int } d^3 x \Bigl [ 2 \beta \partial^0 \bar\beta - 2 \bar \beta 
\partial^0 \beta + (\partial^0 \bar C^i - \partial^i \bar C^0) C_i \nonumber\\
&-& \rho C^0 + \lambda \bar C^0 + (\partial^0 C^i - \partial^i C^0) \bar C_i \Bigr ].
 \end{eqnarray}
These conserved charges
$Q_{(a)b}$ and $Q_g$ obey the following BRST algebra
\begin{eqnarray}
&& Q_b^2 = \frac{1}{2} \{ Q_b, Q_b \} = 0,\; \qquad Q_{ab}^2 =
\frac{1}{2} \{ Q_{ab}, Q_{ab} \} = 0, \nonumber\\
&& Q_b Q_{ab} +
Q_{ab} Q_b \equiv \{Q_b, Q_{ab} \}= 0 \equiv \{ Q_{ab}, Q_b \}, \nonumber\\
&&  i [Q_g, Q_b] = + Q_b,\; \qquad i [Q_g, Q_{ab}] = - Q_{ab}.
\end{eqnarray}
The above algebra plays a key role in the cohomological description of the states of the 
quantum gauge theory in the quantum Hilbert space (QHS).

The algebra in (21) can be derived by exploiting the infinitesimal transformations $s_{(a)b}$ and
$s_g$ and the expressions for $Q_{(a)b}$ and $Q_g$. These are
\begin{eqnarray}
&& s_b Q_b = - i \{ Q_b, Q_b \} = 0,\; \qquad s_{ab} Q_{ab} = - i \{Q_{ab}, Q_{ab} \} = 0, \nonumber\\
&& s_b Q_{ab} = - i \{ Q_{ab},  Q_b \} = 0,\; \qquad
s_{ab} Q_b = - i \{ Q_b, Q_{ab} \} = 0, \nonumber\\ && s_{g} Q_{ab} = - i [Q_{ab}, Q_{g}] = - Q_{ab},
 \qquad s_g Q_{b} = - i [ Q_{b},  Q_g ] = Q_b, \nonumber\\
&& s_{b} Q_{g} = - i [Q_{g}, Q_{b}] = - Q_{b},\;
\qquad s_{ab} Q_{g} = - i [ Q_{g},  Q_{ab} ] = Q_{ab}.
\end{eqnarray}
In the above computations, the factors of $\pm 2$ and $\pm 1$ present in the ghost transformations
(18), play a very crucial role. 
Furthermore, some of the computations in the above are really
non-trivial and algebraically more involved. In particular, in the proof of
$\{Q_b, Q_{ab} \} \equiv \{Q_{ab}, Q_b \} = 0$, one has to exploit the restriction (5) and equations of motion.

The physical state of the QHS is defined as $Q_{(a)b} |phys> = 0$. This condition comes out to be consistent with the Dirac's prescription for the quantization of theories with
first-class constraints [1,2]. The details of the constraints analysis has been performed
in our earlier work [4] where it has been shown that the constrained field equation (5)
can be incorporated in the physicality condition $Q_{(a)b} |phys> = 0$ in a subtle manner
(see, e.g. [4] for details). For our present
Abelian 2-form gauge theory, the BRST and anti-BRST charges play their separate and
independent roles as has been established in [4] by performing a detailed constraint 
analyses of this theory.

\section{Conclusions}

In our present investigation, we have concentrated on the appropriate Lagrangian densities
of the 4D free Abelain 2-form gauge theory that (i) respect the off-shell nilpotent and 
absolutely anticommuting (anti-)BRST symmetry transformations that were derived in our earlier
works [3-5], (ii) are free of a specific Lagrange multiplier 4D vector field which was introduced
in our earlier endeavours to incorporate the analogue of the CF type 
restriction [3-5], (iii) are endowed with terms that carry standard meaning of the
quantum field theory\footnote{ It will be noted that, in our previous attempts [3-5], the kinetic 
term for the massless scalar field turned out to possess a negative sign due
to the constraint field equation.}, and (iv) can be generalized
so as to prove that the present 4D theory is a field theoretic model for the Hodge theory [14,15].

It is pertinent to point out that the Lagrangian densities in (6) and (7) can be recast
into different simple and beautiful forms as is evident from equations (12), (13), (16) and (17).
This should be contrasted, however, with the Lagrangian densities (1) and (4) which cannot be
recast into the above beautiful forms because of (i) the Lagrange multiplier term
(i.e. $L^\mu (B_\mu - \bar B_\mu - \partial_\mu \phi)$), and (ii) the kinetic term
for the massless scalar field (i.e. $- (1/2) \partial_\mu \phi \partial^\mu \phi$).
Thus, it is clear that the Lagrangian densities (6) and (7), that respect the same symmetry
transformations as (1) and (4), are more appealing and more economical than their
counterparts in (1) and (4).

The anticommutativity property of
the nilpotent (anti-)BRST symmetry transformations owes its origin to the
analogue of the CF condition (cf. (5), (11))  which describes a constrained surface on the 4D spacetime manifold. The key insight, for the existence of this relation, comes from
the superfield approach to BRST formalism in the context of our present theory [5]. It is very interesting to note that, despite our present 4D gauge
theory being an {\it Abelian} 2-form gauge theory, an
analogue of the CF condition (which is the hallmark of a {\it non-Abelian} 1-form gauge theory) exists for the sanctity of the anticommutativity property of the
(anti-)BRST symmetry transformations. Recently, we have been able to show the time-evolution invariance
of this restriction in the Hamiltonian formalism [16].

There are a few relevant points that have to be emphasized. First, unlike non-Abelian
1-form theory [13], the above CF type restriction does not connect the auxiliary 
vector fields $B_\mu$ and $\bar B_\mu$ with any kind of (anti-)ghost fields of the theory. 
Rather, the above condition (5) is a relationship between the 
auxiliary fields and scalar field of the theory which are all bosonic in nature. Second, the 
analogue of the CF restriction present in our
Abelian 2-form gauge theory has been shown [3] to have deep connection with the concept of gerbes.
These geometrical objects, at the moment, are one of the very active areas of research in theoretical 
high energy physics. Finally, it would be nice to establish connection
between the above fermionic (anti-)BRST charges and the twisted supercharges 
of the extended supersymmetry algebra. We plan to pursue the above cited issues further 
for our future investigations in the realm of 2-form and higher-form 
(non-)Abelian gauge theories [17].

\noindent
{\bf Acknowledgements:} Financial support from DST, Government of India,  
under the SERC project: SR/S2/HEP-23/2006, is gratefully acknowledged.

\end{document}